# Origin of optical losses in gallium arsenide disk whispering gallery resonators

David Parrain,[1] Christophe Baker,[1] Guillaume Wang,[1] Biswarup Guha,[1] Eduardo Gil Santos,[1] Aristide Lemaitre,[2] Pascale Senellart,[2] Giuseppe Leo,[1] Sara Ducci,[1] and Ivan Favero[1,*]

[1]*Matériaux et Phénomènes Quantiques, Université Paris Diderot, CNRS UMR 7162, Sorbonne Paris Cité, 10 rue Alice Domon et Léonie Duquet, 75013 Paris, France*
[2]*Laboratoire de Photonique et de de Nanostructures, CNRS UPR 20, Route de Nozay, 91460 Marcoussis, France*
[*]*ivan.favero@univ-paris-diderot.fr*

**Abstract:** Whispering gallery modes in GaAs disk resonators reach half a million of optical quality factor. These high Qs remain still well below the ultimate design limit set by bending losses. Here we investigate the origin of residual optical dissipation in these devices. A Transmission Electron Microscope analysis is combined with an improved Volume Current Method to precisely quantify optical scattering losses by roughness and waviness of the structures, and gauge their importance relative to intrinsic material and radiation losses. The analysis also provides a qualitative description of the surface reconstruction layer, whose optical absorption is then revealed by comparing spectroscopy experiments in air and in different liquids. Other linear and nonlinear optical loss channels in the disks are evaluated likewise. Routes are given to further improve the performances of these miniature GaAs cavities.

## References and links

1. B. Gayral, J. M. Gérard, A. Lemaître, C. Dupuis, L. Manin, and J. L. Pelouard, "High-Q wet-etched gaas microdisks containing inas quantum boxes," Appl. Phys. Lett **75**, 1908 (1999).
2. A. Kiraz, P. Michler, C. Becher, B. Gayral, A. Imamoglu, L. Zhang, E. Hu, W. V. Schoenfeld, and P. M. Petroff, "Cavity-quantum electrodynamics using a single inas quantum dot in a microdisk structure," Appl. Phys. Lett. **78**, 3932 (2001).
3. E. Peter, P. Senellart, D. Martrou, A. Lemaître, J. Hours, J. M. Gérard, and J. Bloch, "Exciton-photon strong coupling regime for a singe quantum dot embedded in a microcavity," Phys. Rev. Lett. **96**, 067401 (2005).
4. A. Andronico, I. Favero, and G. Leo, "Difference frequency generation in gaas microdisks," Opt. Lett. **33**, 2026-2028 (2008).
5. P. Kuo, J. Bravo-Abad, and G. Solomon, "Second-harmonic generation using quasi-phasematching in a gaas whispering-gallery-mode cavity," Nat. Commun. **5**, 3109 (2014).
6. S. Mariani, A. Andronico, A. Lemaître, I. Favero, S. Ducci, and G. Leo, "Second-harmonic generation in algaas microdisks in the telecom range," Opt. Lett. **39**, 3062-3065 (2014).
7. S. L. McCall, A. F. J. Levi, R. E. Slusher, S. J. Pearton, and R. A. Logan, "Whispering-gallery mode microdisk lasers," Appl. Phys. Lett. **60**, 289 (1992).
8. T. Ide, T. Baba, J. Tatebavashi, S. Iwamoto, T. Nakaoka, and Y. Arakawa, "Room temperature continuous wave lasing in inas quantum-dot microdisks with air cladding," Opt. Express **13**, 1615-1620 (2005).
9. S. Reitzenstein, A. Bazhenov, A. Gorbunov, C. Hofmann, S. Münch, A. L. Löffler, M. Kamp, J. P. Reithmaier, V. D. Kulakovskii, and A. Forchel, "Lasing in high-Q quantum-dot micropillar cavities,"Appl. Phys. Lett. **89**, 051107 (2006).
10. L. Ding, C. Baker, P. Senellart, A. Lemaître, S. Ducci, G. Leo, and I. Favero, "High frequency gaas nano-optomechanicsl disk resonator," Phys. Rev. Lett. **105**, 263903 (2010).

11. L. Ding, C. Baker, P. Senellart, A. Lemaître, S. Ducci, G, Leo, and I. Favero, "Wavelength-sized gaas optomechanical resonators with gigahertz frequency," Appl. Phys. Lett. **98**, 113108 (2011).
12. D. Parrain, C. Baker, T. Verdier, P. Senellart, A. Lemaître, S. Ducci, G. Leo, and I. Favero, "Damping of optomechanical disks resonators vibrating in air," Appl. Phys. Lett. **100**, 242105 (2012).
13. C. P. Michael, K. Srinivasan, T. J. Johnson, O. Painter, K. H. Lee, K. Hennessy, H. Kim, and E. Hu, "Wavelength- and material-dependent absorption in gaas and algaas microcavities," Appl. Phys. Lett. **90**, 051108 (2007).
14. J. C. L. Ding, C. Baker, A. Andronico, D. Parrain, P. Senellart, A. Lemaître, S. Ducci, G. Leo, and I. Favero, "Gallium arsenide disk optomechanical resonators," in *Handbook of Optical Microcavities* (PanStanford, 2014).
15. I. Favero, "Gallium Arsenide disks as optomechanical resonators," in *Cavity Optomechanics* (Springer, 2014).
16. C. Baker, "On-chip nano-optomechanical whispering gallery resonators," PhD thesis Université Paris Diderot (2013).
17. E. Kuramochi, H. Taniyama, T. Tanabe, A. Shinya, and M. Notomi, "Ultrahigh-Q two-dimensional photonic crystal slab nanocavities in very thin barriers," Appl. Phys. Lett. **93**, 111112 (2008).
18. S. Combrié, A. De Rossi, Q. V. Tran, and H. Benisty, "Gaas photonic crystal cavity with ultrahigh Q: microwatt non-linearity at 1.55 μm," Opt. Lett. **33**, 1908-1910 (2009).
19. P. B. Deotare, M. W. McCutcheon, I. W. Frank, M. Khan, and M. Loncar, "High quality factor photonic crystal nanobeam cavities," Appl. Phys. Lett. **94**, 121106 (2009).
20. Y. Taguchi, Y. Takahashi, Y. Sato, T. Asano, and S. Noda, "statistical studies of photonic heterostructure nanocavities with an average Q factor of three million," Opt. Express **19**, 11916-11921 (2011).
21. A. Andronico, "Etude électromagnétique d'émetteurs intégrés infrarouges et terahertz en AlGaAs," PhD thesis Université Paris Diderot (2008).
22. A. Andronico, X. Caillet, I. Favero, S. Ducci, V. Berger and G. Leo, "Semiconductor microcavities for enhanced nonlinear optics interactions," J. Eur. Opt. Soc, Rapid Publ. **3**, 08030 (2008).
23. L. Ding, P. Senellart, A. Lemaitre, S. Ducci, G. Leo, and I. Favero, "GaAs micro-nanodisks probed by a looped fiber taper for optomechanics applications," Proc. SPIE **7712**, 771211 (2010).
24. M. Skorobocgatiy, G. Bégin, and A. Talneau, "Statistical analysis of geometrical imperfections from the images of 2d photonic crystals," Opt. Express **13**, 2487-2502 (2005).
25. C. G. Poulton, C. Koos, M. Fujii, A. Pfrang, T. Schimmel, J. Leuthold, and W. Freude, "Radiation modes and roughness loss in high index-contrast waveguides," IEEE J. Sel. Top. Quantum Electron. **12**, 1306-1321 (2006).
26. M. Kuznetsov, and H. Haus, "Radiation loss in dielectric waveguide structures by the volume current method," IEEE J. Quant. Electron. **19**, 1505-1514 (1983).
27. M. Borselli, K. Srinivasan, P. E. Barclay, and O. Painter, "Rayleigh scattering, mode coupling, and optical loss in silicon microdiks," App. Phys. Lett. **85**, 3693 (2004).
28. T. Barwicz, and H. Haus, "Three-dimensional analysis of scattering losses due to sidewall roughness in microphotonic waveguides," J. Lightwave Technol. **23**, 2719- (2005).
29. S. G. Johnson, M. Ibanescu, M. Skorobogatiy, O. Weisberg, J. Joannopoulos, and Y. Fink, "Perturbation theory for Maxwell's equations with shifting material boundaries," Phys. Rev. E **65**, 066611- (2002).
30. S. G. Johnson, M. Povinelli, M. Soljacic, A. Karalis, S. Jacobs, and J. Joannopoulos, "Roughness losses and volume-current methods in photonic crystal waveguides," Appl. Phys. B **81**, 283-293 (2005).
31. N. Hill, "Integral-equation perturbative approach to optical scattering from rough surfaces," Phys. Rev. B **24**, 7112- (1981).
32. J. E. Heener, T. C. Bond, and J. S. Kallman, "Generalized formulation for performance degradations due to bending and edge scattering loss in microdisk resonators," Opt. Express **15**, 4452-4458 (2007).
33. T. Carmon, L. Yang, and K. J. Vahala, "Dynamical thermal behavior and thermal self-stability of microcavities," Opt. Express **12**, 4742-4750 (2004).
34. V. R. Almeida, and M. Lipson, "Optical bistability on a silicon chip," Opt. Lett. **29**, 2387-2389 (2004).
35. D. A. Kleinman, R. C. Miller, and W. A. Nordland, "Two-photon absorption of nd laser radiation in gaas," Appl. Phys. Lett. **23**, 243-244 (1973).
36. B. Bosacchi, J. Bessey, and F. Jain, "Two-photon absorption of neodymium laser radiation in gallium arsenide," J. Appl. Phys. **49**, 4609 (1978).
37. W. C. Hurlbut, Y. S. Lee, K. L. Vodopyanov, P. S. Kuo, and M. M. Fejer, "Multiphoton absorption and nonlinear refraction of gaas in the mid-infrared," Opt. Lett. **32**, 668-670 (2007).
38. S. Krishnamurthy, Z. G. Yu, L. P. Gonzalez, and S. Guha, "Temperature- and wavelength-dependent two-photon and free-carrier absorption in gaas, inp, gainas, and inasp," J. Appl. Phys. **109**, 033102 (2011).
39. A. Baca, and C. Ashby, *Fabrication of GaAs devices* (The Institution of Engineering and Technology, 2005).
40. C. Baker, C. Belacel, A. Andronico, P. Senellart, A. Lemaitre, E. Galopin, S. Ducci, G. Leo, and I. Favero, "Critical optical coupling between a GaAs disk and a nanowaveguide suspended on the chip," Appl. Phys. Lett. **99**, 151117 (2011).
41. E. McLaughlin, "The thermal conductivity of liquids and dense gases," Chem. Rev. **64**, 389-428 (1964).
42. G. M. Hale, and M. R. Querry, "Optical constants of water in the 200nm to 200μm wavelength region," Appl. Opt. **12**, 555-563 (1973).

43. H. Oigawa, J. J. Fan, Y. Nannichi, H. Sugahara, and M. Oshima,"Universal passivation effect of (nh4)2 sx treatment on the surface of iii-v compound semiconductors," Jpn. J. Appl. Phys. **30**, L322 (1991).
44. C. I. H. Ashby, K. R. Zavadil, A. G. Baca, P-C. Chang, B. E. Hammons, and M. J. Hafich, "Metal-sulfur-based air-stable passivation of gaas with very low surface densities," Appl. Phys. Lett. **76**, 327 (2000).
45. E. Yablonovitch, C. J. Sandroff, R. Bhat, and T. Gmitter, "Nearly ideal electronic properties of sulfide coated gaas surfaces," Appl. Phys. Lett. **51**, 439 (1987).
46. G. Mariani, R. B. Laghumavarapu, C. Tremolet de Villers, J. Shapiro, P. Senanayake, A. Lin, B. J. Schwartz, and D. Huffaker, "Hybrid conjugated polymer solar cells using patterned GaAs nanopillars," Appl. Phys. Lett. **97**, 013107 (2010).
47. S. D. Offsey, J. M. Woodall, A. C. Warren, P. D. Kirchner, T. I. Chappel, and G. D. Pettit, "Unpinned (100) gaas surfaces in air using photochemistry," Appl. Phys. Lett. **48**, 475 (1986).
48. V. Berkovits, D. Paget, A. Karpenko, V. Ulin, and O. Tereshchenko,"Soft nitridation of gaas (100) by hydrazine sulfide solutions: Effect on surface recombination and surface barrier," Appl. Phys. Lett. **90**, 022104 (2007).
49. V. Berkovits, V .Ulin, O. Tereshchenko, D. Paget, A. Rowe, P. Chiaradia, B. Doyle, and S. Nannarone, "Chemistry of wet treatment of GaAs (111) b and GaAs (111) a in hydrazine-sulfide solutions," J. Electrochem. Soc. **158**, D127-D135 (2011).

**1. Introduction**

Whispering Gallery Mode (WGM) Gallium Arsenide (GaAs) optical cavities incorporate the beneficial optical properties of the GaAs material into an optical mode of ultra-small volume V and high quality factor Q. They have led to important applications in several contexts where enhanced light-matter interaction is required like cavity-QED [1-3], nonlinear optics [4-6], low-threshold laser devices [7-9] and optomechanics [10-12]. Even if based on somewhat distinct concepts, all these applications generally benefit from large values of Q/V to enhance the local electromagnetic energy density.

For optical wavelengths in the near infrared, the mode volume of GaAs disks WGMs can be made sub-$\mu m^3$ [3,11], while Q reaches $5.10^5$ in the best technological realizations, be they based on wet [13,14] or dry etching fabrication [15,16]. While this brings these resonators close to the state of the art of Q/V in semiconductors [17-20], they could in principle perform even better. Indeed the optical Q expected from bending losses of a disk can be made ultra-high. In a disk of radius 1 µm and thickness 300 nm, enabling sub-$\mu m^3$ WGM volume, the design Q is for example larger than $10^{10}$ at a wavelength of 1 µm, in the transparency region of GaAs. In this view, it appears crucial to understand what currently experimentally precludes GaAs cavities to reach these ultimate light confinement performances.

Here we address these questions by a high-resolution Transmission Electron Microscope (TEM) investigation of GaAs disk resonators and by systematic optical experiments performed on their WGMs in linear and non-linear regimes. The TEM analysis provides precise structural information about the resonators and their surfaces, which permits a quantitative assessment of optical scattering losses by means of an improved Volume Current Method. In high-Q GaAs WGMs, scattering by imperfections appears to be eclipsed by a residual level of optical absorption. This finding is supported by optical experiments at high optical power, where nonlinear thermo-optic and two-photon effects are carefully analyzed to gain insight in optical absorption processes. Finally, comparative WGM spectroscopy experiments are performed both in air and in liquid, which reveal the important role played by surface absorption. A consistent picture of optical dissipation in GaAs disks eventually emerges, which points towards routes to best exploit their potential.

## 2. Bending losses

In this section, we first discuss radiative losses in GaAs disks resulting from their curvature. The material absorption is assumed to be negligible for the wavelengths of interest, far enough from the gap. This assumption will be relaxed in sections 6 and 7. For consistency we start with a rapid description of WGMs and their computation techniques. Thanks to the rotation invariance of a disk resonator around its axis Z and to the isotropy of the linear dielectric response of the material, the Maxwell equations in this geometry can be solved by employing an azimuthal dependence $\exp(im\theta)$ of the electromagnetic field. This allows reducing the 3-dimensional electromagnetic problem of a disk to a 2-dimensional one, which is then treated by approximate semi-analytical means like the Effective Index Method (EIM) or by exact numerical techniques like FEM. In the EIM, the field is usually assumed to be quasi-TE (-TM) meaning that the only non-vanishing component of the magnetic (electric) field is orthogonal to the disk plane, noted $F_z$. Combined with an assumption of separation of the radial (r) and vertical (z) variables in any plane containing the Z-axis, the quasi-TE (TM) approach leads to solutions for the field component $F_z$ with a radial dependence of the form $J_m(k_p n_{eff} r)$ in the disk. Here $J_m$ is the first kind Bessel function of order m, $n_{eff}$ the effective index of the TE or TM mode in a GaAs slab of same thickness as the disk, and $k_p$ the radial wave number [21,22]. p gives the number of lobes of $F_z$ along the radial direction. In the EIM, WGMs are hence naturally classified with a radial and an azimuthal integer p and m, but this classification remains useful in numerical computations as well even if these do not rely on EIM assumptions. For example Fig. 1 shows a 2D axisymmetric FEM simulation in the (r,z) plane of a WGM of a disk of thickness 200 nm, for which the EIM assumptions start to break down. Still the dominant electric field component $E_r$, which is proportional to the magnetic field $H_z$, follows closely a Bessel-like behavior with p=4, and the mode is identified as being TE (p=4,m=11).

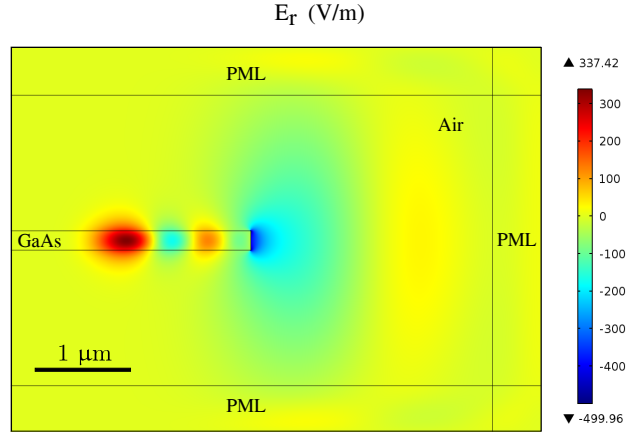

Fig. 1. Axisymmetric FEM computation of the TE (p=4, m=11) WGM of a GaAs disk of radius 2.5 μm and thickness 200 nm. The cross section of the disk is visualized. The vertical axis is Z, while the horizontal axis corresponds to the radial direction. The dominant radial component of the electric field E is shown in arbitrary units, together with PMLs employed for the computation of $Q_{rad}$.

The prediction of the radiation loss rate $\kappa_{rad}$ of a given WGM of angular frequency $\omega_{cav}$ requires a very accurate description of the mode's spatial pattern. For this reason, the approximate EIM approach is often quantitatively wrong in computing the radiative quality factor $Q_{rad}=\omega_{cav}/\kappa_{rad}$ and fully numerical techniques are more indicated. In Fig. 1, Perfectly Matched Layers (PMLs) are shown that emulate far-field absorption of the radiated field in

FEM with the Comsol software. For comparison purposes, we also employed axisymmetric Finite Difference Frequency Domain homemade codes to compute $Q_{rad}$ and found very solid agreement with FEM results. Figure 2 plots the calculated values of $Q_{rad}$ for WGMs of 200 nm thick GaAs disks of radius 5 and 2.5 µm. The values are given for modes in the wavelength range 1500-1600 nm, in the transparency region, which we employed in past experiments [10,14,16]. TM modes exhibit moderate $Q_{rad}$ of at best a few thousands, when TE modes can reach $Q_{rad}$ in excess of $10^9$. We estimate that our numerical computation saturates in precision around $10^{10}$ hence we did not plot explicitly points beyond that value. For the disk of radius 2.5 µm for example, we infer simply that $Q_{rad}$ exceeds $10^{10}$ for the TE (p=1) mode. For the disk of radius 5 µm, $Q_{rad} > 10^8$ for TE modes of $p \leq 5$. These modes are easily mode-matched to optical fiber tapers and hence accessible in experiments [23]. In these experiments, in contrast to the above radiative predictions, the best measured optical Qs saturate at half a million [14]. It is the purpose of the next sections to precisely establish which factors contribute to this difference.

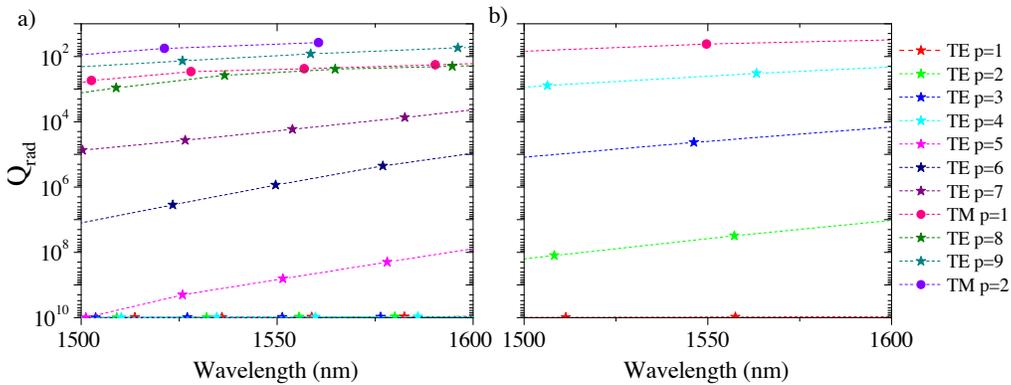

Fig. 2. WGM radiative quality factor $Q_{rad}$ set by bending losses of GaAs disks. (a) Disk radius of 5 µm and thickness of 200 nm. (b) Disk radius of 2.5 µm and thickness of 200 nm. The values are given for TE and TM-like modes of varying radial p number, in the 1.55 µm wavelength range.

## 3. Fabrication of GaAs disks

The GaAs disks employed in this work are fabricated out of an extra-pure wafer consisting of a layer of GaAs (200 nm) and of $Al_{0.8}Ga_{0.2}As$ (1.8 µm thick) grown by Molecular Beam Epitaxy (MBE) on a semi-insulating GaAs substrate. Using e-beam lithography with a negative resist (MaN 2403 or 2401), disks are patterned on the surface and then wet-etched in a two-step process [14]. The first non-selective wet etch relies on a 1:1:1 mixture of hydrobromic acid, potassium dichromate and acetic acid, and is carried-out at 4°C. It gives rise to a GaAs/AlGaAs pillar-like structure, whose circular shape depends crucially on stirring conditions in the etching solution and is improved by keeping the sample at rest. In a second step, also carried-out at 4°C, a 1:20 diluted hydrofluoric (HF) acid solution selectively under-etches the $Al_{0.8}Ga_{0.2}As$ layer to form the disk pedestal while leaving the GaAs disk unaffected. During this step, regular dips in a KOH solution are employed to remove AlGaAs etch byproducts from the sample surface. Figure 3 shows SEM images of GaAs disks fabricated following this process, with a radius that typically varies from 0.5 to 25 µm depending on experiments. With this SEM inspection, the geometrical imperfections of the disk are barely resolvable. Figure 3(a) shows a fabricated disk that looks very regular in the shape, while Fig.

3(b) displays clean surfaces that are virtually free of defects or residual roughness. Figure 3(c), with the strongest zoom, provides a hint of residual surface imperfections, at a level that the SEM cannot properly analyze.

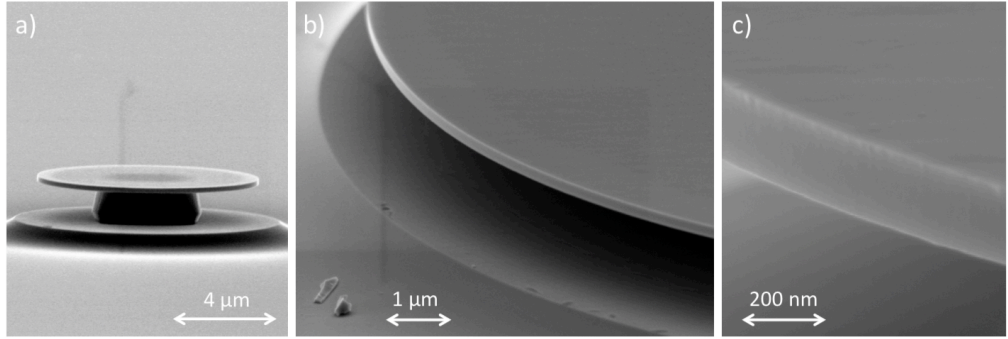

Fig. 3. SEM images of fabricated GaAs disks. (a) Complete disk of diameter 8 μm and thickness 200 nm. (b) Zoom on the top-surface and sidewall of a large GaAs disk of thickness 200 nm, with no residual roughness visible at that scale. (c) Close-up on a GaAs disk sidewall showing barely resolvable residual imperfections at the finest scale achievable in the SEM.

## 4. TEM analysis

This section focuses on the departure of GaAs resonators from an ideal circular geometry. This departure includes both residual roughness of the disk sidewalls and non-perfectly circular geometry of the disk. These geometrical imperfections are analyzed by ultra-high resolution TEM experiments, to circumvent the limited resolution of the SEM. We employ a TEM apparatus corrected from spherical aberrations, which reaches a spatial resolution of 80 pm (Jeol ARM 200F cold FEG). The best contrast in TEM images on GaAs crystal is obtained for a sample thickness between 50 and 200 nm. While we fabricated 50 nm thick samples showing excellent contrast, we will focus here on results obtained with 200 nm, which is closer to the thickness of disks employed in optical experiments. The disks are fabricated as described above. Next, the finished sample is turned upside-down over a TEM copper grid and pressed and rubbed against it. This crude technique detaches about 1% of the disks from their pedestal and deposits them on the carbon film covering the grid. While the technique provides good conditions for TEM imaging, it has a poor yield and the deposited disks are frequently broken. Figure 4 shows an example of a GaAs disk of diameter 5 μm deposited this way and imaged with an increasing magnification factor of the TEM from a) to f). In Fig. 4(a) the disk appears perfectly circular and the pedestal appears in dark. The square shape of the pedestal is induced here by the specific conditions of the HF etching step and plays no role in the analysis below. In the series of zooms from b) to e), the complexity of an interface can be appreciated: starting from a remarkably smooth circle in b), the boundary looses its regularity at smaller scale and it becomes difficult to distinguish a precise border between the ordered GaAs material and its external environment. In d), e) and f), an amorphous boundary zone is progressively revealed between the GaAs crystal and the exterior. Electron Energy Loss Spectroscopy (EELS) experiments were carried-out on this amorphous zone, which revealed the prominent presence of Gallium and Arsenic atoms. However, despite careful checks, no Oxygen atoms were detected by this EELS analysis, contradicting the usual picture of an oxide layer formed on the surface. We infer that the fabrication process produces an amorphous GaAs reconstruction layer at the surface of the disks.

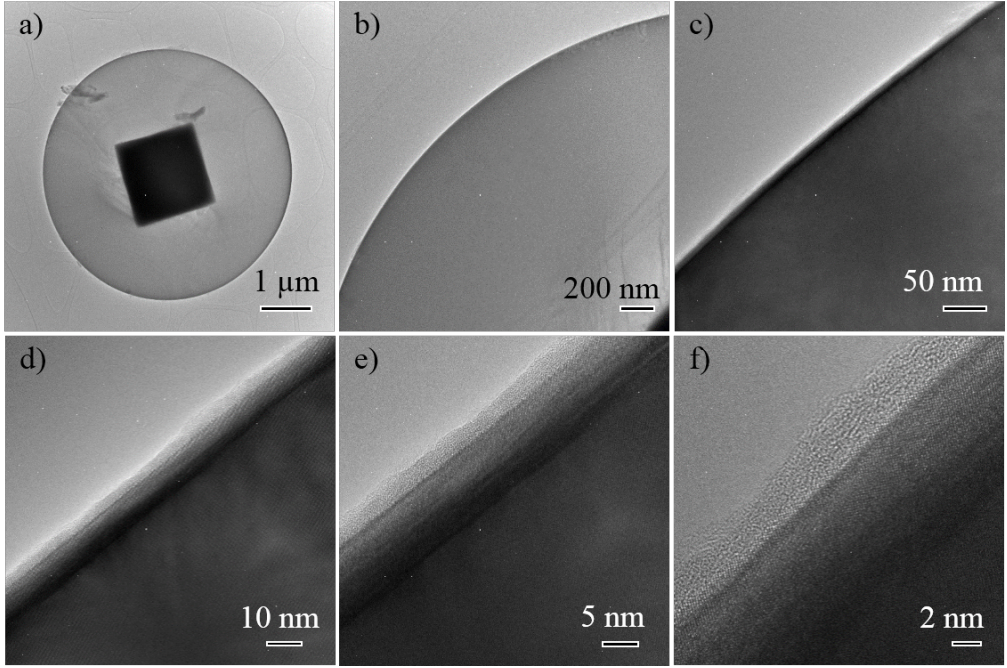

Fig. 4. TEM images of a GaAs disk of radius 2.5 µm and thickness 200 nm. (a) Complete disk with magnification factor of × 3000. (b) × 10 000. (c) × 50 000. (d) × 200 000. (e) ×400 000. (f) × 800 000.

Because of the high quality of the epitaxial material, the upper and lower surfaces of the disk are planar and we assume that geometrical imperfections are concentrated on the disk's curved sidewall. The sidewall outline is projected on a 2D contour of the disk in the plane of TEM observation, and analyzed as sketched in Fig. 5. Starting from a 2D TEM image of the disk (Fig. 5(a)), the contour is extracted using the 'edge' algorithm of Mathlab. This results in a continuous 2D contour (Fig. 5(b)), which is then fitted by a circle defined by its center and its radius R (Fig. 5(c)). The center being known, an azimuthal angle $\theta$ can be employed to parameterize the distance $r(\theta)$ of the contour to the center (Fig. 5(d)). The information on the irregular contour is represented in 1D to facilitate further statistical analysis (Fig. 5(e)). We characterize the disk contour's departure from ideality by the parameter $\delta r(\theta) = r(\theta) - R$, corresponding to the distance to the fit circle of radius R.

This contour analysis requires using the right scale in TEM imaging. Indeed a small magnification factor leads to a reduced spatial resolution that can affect the obtained statistical information. On the other hand, at large magnification factor, the presence of the amorphous surface reconstruction layer renders the definition of an exact boundary between disk and exterior difficult. Additionally, the restricted imaging field of Figs. 4(e) and 4(f) (edge of 50 and 20 nm respectively) induces under-estimation of some statistical quantities needed for optical scattering problems. For this reason, a multi-scale approach would in principle be welcome for the study of contours. Still in this work, we adopted a single imaging scale that permits the observation of the complete disk like shown in Fig. 4(a), for two reasons.

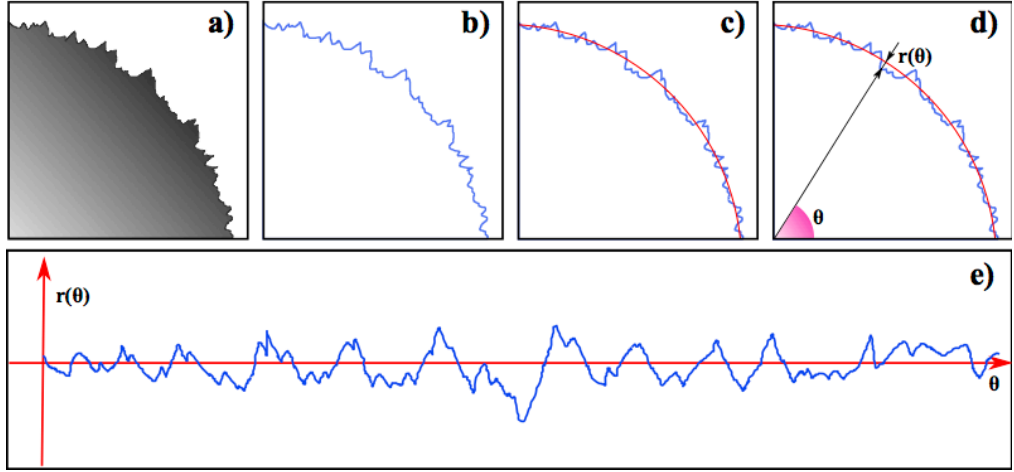

Fig. 5. Principles of disk contour analysis. (a) Image of an irregular disk. (b) A rough contour boundary is extracted. (c) The irregular contour is fitted with a circle arc. (d) The distance to the fit circle is registered as a function of the azimuthal angle θ. (e) The obtained irregular contour is plotted on a horizontal axis for further analysis.

Firstly, it is hard to recover the correct statistical analysis of the whole contour from a series of magnified images of too restricted sub-parts of the contour. Indeed, in our procedure, each image of the series leads to its own fit circle that differs from the fit circle of the global contour, creating a bias in the analysis when the stitched information about the whole contour is needed. Hence the complete disk should best be visible on a single TEM image. Secondly, this latter imaging scale provides sufficient resolution to resolve the contour irregularities needed for the analysis. In order to verify this, we simulated the effect of images with a limited spatial resolution on our contour extraction protocol. This was done by convoluting a Gaussian function with randomly generated 2D disks (to artificially mimic blurry/unresolved images) and applying our contour extraction protocol. Increasing the Gaussian width progressively affected δr(θ) and its auto-correlation function, which adopted a Gaussian shape. Under real experimental conditions with the resolution of our TEM apparatus, this effect turned out to be negligible. With the same convolution procedure, we also simulated the effect of the image pixel size, which was to create a Dirac-like peak at the origin in the auto-correlation function of δr(θ). Since this peak is deterministically associated to the pixel size, it can be disregarded in the final data. For all these reasons, in the results shown below, the imaging scale is fixed to image a whole disk and the pixel-effect is not further considered.

Now that a proper TEM magnification factor is chosen, the question of multiple scales in the analysis of contour irregularities can be dealt with as follows. At large scale, the irregularities may present some level order. For example, if a disk is slightly elliptical or if the contour oscillates in a regular manner, it may be approached by a series of sinusoidal functions. Once this "wavy" part of the contour is removed, a non-ordered residue Res(θ) associated to the residual roughness remains, whose correlation function must decay to zero at large distance. The distinction between waviness and roughness is formally written like [24]:

$$\delta r(\theta) = r(\theta) - R = \sum_{m=1}^{N} A_m \cos(m\theta + \phi_m) + Res(\theta)$$

Note that this distinction between waviness and roughness is somewhat arbitrary since it depends on the imaging scale. Mathematically, the contour r(θ) is 2π-periodic and can be approached arbitrarily close by a Fourier series. In practice, the imaging scale and the level of resolution determine at which order N the observer sets the frontier between waviness and roughness. In what follows N is the smallest integer that leads to an auto-correlation function of Res(θ) decaying at large distance below 20% of its peak value.

Figure 6 illustrates this statistical approach by showing the results obtained on the complete disk of Fig. 4(a). Figure 6(a) plots the contour extracted from the TEM image in blue, together with the waviness in red, corresponding to the fit function (in nm):

$$r(\theta) = 2457 - 1.35\cos(\theta - 0.60) - 12\cos(2\theta + 0.36)$$
$$+ 17\cos(3\theta + 1.34) + 3.39\cos(4\theta + 0.25) - 0.9\cos(5\theta + 0.03) + 1.4\cos(6\theta + 0.23)$$
$$+ 0.6\cos(7\theta - 0.06) - 1.1\cos(8\theta + 0.09) + Res(\theta)$$

Figure 6(b) shows the spatial auto-correlation function of the waviness, whose amplitude oscillates without decaying far from the origin. Figure 6(c) shows the residue Res(θ), which fluctuates essentially between + and – 3 nm with no apparent spatial order. Figure 6(d) is the auto-correlation of the residue, which decays to zero far from the origin. It can be fitted by an exponential decay or by the sum of a sine wave and an exponential [25] to reduce the fit error. The parameters of the exponential fit provide effective values for the correlation length $L_c$ of the roughness and its amplitude σ ($σ^2$ is the auto-correlation value at the origin).

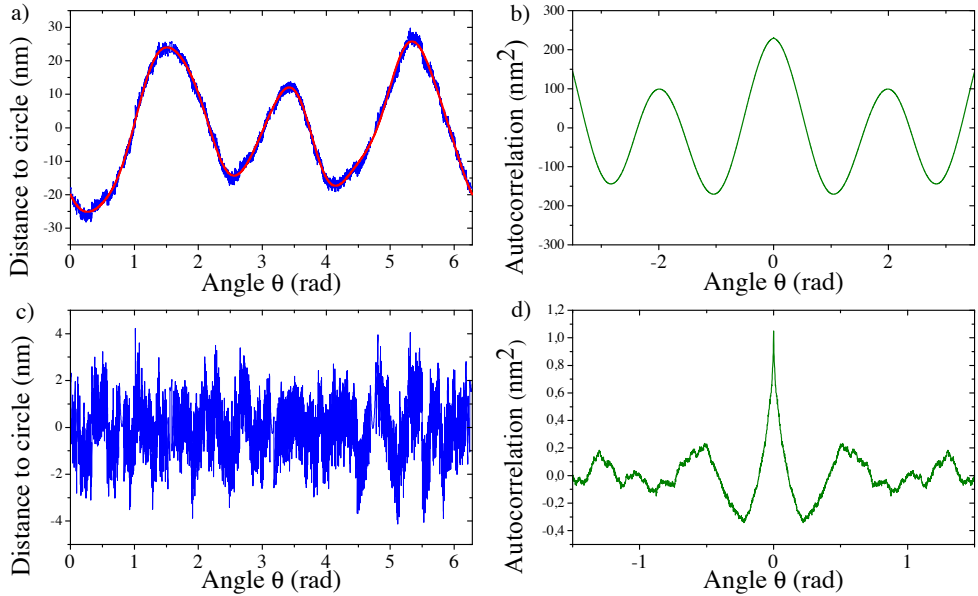

Fig. 6. Contour analysis of a complete disk. (a) Azimuthal representation of the distance δr(θ) to the fit disk. Data are in blue and the fitted waviness function is in red. (b) Auto-correlation function of the contour r(θ). (c) Residue of the contour once the waviness is subtracted (d) Auto-correlation function of the residue.

While this first example of analysis was performed on a complete disk, we also extended the analysis to some partially broken disks where one half to two thirds of the contour remained intact after deposition on the TEM grid. To this purpose, we first extrapolated each partial contour to a periodic function with period close to 2π, and employed the above statistical procedure. We checked the consistency of this extrapolation method by analyzing quarter sections of the complete disk and comparing the outcomes of the statistical analysis with those obtained on the complete contour. The extrapolation gave good agreement for disk sections of quarter length or more, and finally allowed obtaining an analysis of the roughness of four other disks, whose parameters are summarized in Table 1. There is a quantitative resemblance between the 5 analyzed disks in total, with the amplitude of residual roughness lying between 0.5 and 1.5 nm and the correlation length between 20 and 80 nm.

Table 1. Analysis of the residual roughness of five GaAs disks having distinct contour angular extent.

|   | disk 1 (complete disk $2\pi$) | disk 2 ($\pi/2$) | disk 3 ($4\pi/3$) | disk 4 ($5\pi/4$) | disk 5 ($\pi/2$) |
|---|---|---|---|---|---|
| Radius R (nm) | 2457 | 2400 | 3014 | 3169 | 3188 |
| Amplitude $\sigma$ (nm) | 1.05 | 0.57 | 1.41 | 0.57 | 1.15 |
| $L_c$ (nm) | 68 | 22 | 68 | 27 | 64 |

## 5. Optical scattering losses

In order to assess the impact of the surface irregularities on WGM photon scattering, we employ a perturbative approach of Maxwell's equations, the so-called Volume Current Method (VCM) [26-28], in the limit of a vanishing material absorption. This perturbative treatment is justified by the small amplitude of contour irregularities (30 nm waviness and 1 nm residual roughness) as compared to the typical disk radius of a few microns. The unperturbed electric field $\mathbf{E}^0(\mathbf{r})$ (associated to a perfectly circular contour) induces a polarization current $\mathbf{J}(\mathbf{r})$ in the geometrically perturbed volume of the disk (the irregular contour) that radiates an electric field $\delta\mathbf{E}(\mathbf{r})$ such that the solution of the perturbed problem becomes $\mathbf{E}^0(\mathbf{r}) + \delta\mathbf{E}(\mathbf{r})$. This total field must obey the equation:

$$\nabla \wedge (\nabla \wedge \mathbf{E}(r)) - \frac{\omega^2}{c^2}\varepsilon_r^0(\mathbf{r})\mathbf{E}(r) = \frac{\omega^2}{c^2}\delta\varepsilon_r(\mathbf{r})\mathbf{E}(r)$$

with $\delta\varepsilon_r(\mathbf{r})=(n^2-1)[\Theta(r-R)-\Theta(r-R-\delta r(\theta))]=(n^2-1)\delta(r-R)\delta r(\theta)$, where the Dirac function $\delta(r)$ is the derivative of the Heaviside function $\Theta(r)$, and where n is the refractive index of the disk resonator material. The above equation can be rewritten as:

$$\nabla \wedge (\nabla \wedge \delta\mathbf{E}(r)) - \frac{\omega^2}{c^2}\varepsilon_r^0(\mathbf{r})\delta\mathbf{E}(r) = \frac{\omega^2}{c^2}\delta\varepsilon_r(\mathbf{r})\mathbf{E}^0(r) + \frac{\omega^2}{c^2}\delta\varepsilon_r(\mathbf{r})\delta\mathbf{E}(r)$$

where it clearly appears that the weak field $\delta\mathbf{E}(\mathbf{r})$ is driven by the right hand-side source. In the vast majority of the VCM literature the second-order term of the source is assumed to be negligible and the perturbation polarization current becomes $\mathbf{J}(\mathbf{r})=-i\omega\varepsilon_0\delta\varepsilon_r(\mathbf{r})\mathbf{E}^0(\mathbf{r})$. However, the discontinuity of the orthogonal component of the electric field at the interface of two dielectrics can lead to a breakdown of the usual VCM formulation. This is notably important in TE WGMs where the electric field has an appreciable component orthogonal to the disk's sidewall. This problem in the perturbation of Maxwell's equations was solved in recent works [29,30] following early discussions [31]. In the spirit of these recent developments, we employ here a corrected expression for the polarization current [32]:

$$\mathbf{J}(\mathbf{r}) = -i\omega\big[\varepsilon_0\Delta\varepsilon_r\mathbf{E}^0_\parallel(\mathbf{r}) - \Delta\varepsilon_r^{-1}\mathbf{D}^0_\perp(\mathbf{r})\big]\delta(r-R)\delta r(\theta)$$

with $\Delta\varepsilon_r = n^2-1$ and $\Delta\varepsilon_r^{-1} = 1/n^2 - 1$. In this expression $\mathbf{E}_{||}$ and $\mathbf{E}_\perp$ are the field components parallel and orthogonal to the surface and $\mathbf{D}$ is the displacement vector, such that $\mathbf{J}$ is continuous at the interface. From the polarization current $\mathbf{J}$, the potential vector $\mathbf{A(r)}$ associated to the correction field $\delta\mathbf{E(r)}$ is expressed in form of a Green function:

$$\mathbf{A(r)} = \frac{\mu_0}{4\pi}\int_V \mathbf{J(r')}\frac{e^{-i\frac{\omega}{c}|r-r'|}}{|\mathbf{r}-\mathbf{r'}|}d^3\mathbf{r'} \sim \frac{\mu_0}{4\pi}\frac{e^{-i\frac{\omega}{c}r}}{r}\int_V \mathbf{J(r')}\,e^{i\frac{\omega}{c}\mathbf{u_r\cdot r'}}d^3\mathbf{r'}$$

where the second approximate expression holds in the far-field $|r-r'|\gg R$. In the far field, we use the simplified vectorial relation $\nabla\wedge = -i(\omega/c)\,\mathbf{u_r}\wedge$ and express the electromagnetic fields radiated by the perturbation polarization current as:

$$\delta\mathbf{E(r)} = i\,\omega\,\mathbf{u_r}\wedge(\mathbf{u_r}\wedge\mathbf{A}) \quad ; \quad \delta\mathbf{H(r)} = -i\,\omega\sqrt{\frac{\varepsilon_0}{\mu_0}}(\mathbf{u_r}\wedge\mathbf{A})$$

Starting from a given unperturbed WGM field $\mathbf{E^0}$ calculated by FEM, the injection of an arbitrary contour perturbation $\delta r(\theta)$ leads to the total field $\mathbf{E} = \mathbf{E^0} + \delta\mathbf{E}$, which is obtained at each point $\mathbf{r}$ by numerically integrating the FEM results through the above formula. Once $\mathbf{E}$ and $\mathbf{H}$ are known, the radiated electromagnetic power P is computed by summing in the far field the flux of the Poynting vector across a sphere S surrounding the disk $P=\int_S \frac{1}{4}\cdot(\mathbf{E}\wedge\mathbf{H}^*+c.c.)\cdot\mathbf{u_r}\,r^2 d\Omega$. The quality factor $Q = \omega_{cav}/\kappa$ of the WGM, with $\kappa$ the energy decay rate of the mode, is then obtained by dividing the energy stored in the mode by the energy lost during one optical cycle $Q = (\omega/P)\int_V \frac{1}{2}\varepsilon_0|\mathbf{E^0(r)}|^2 dV$.

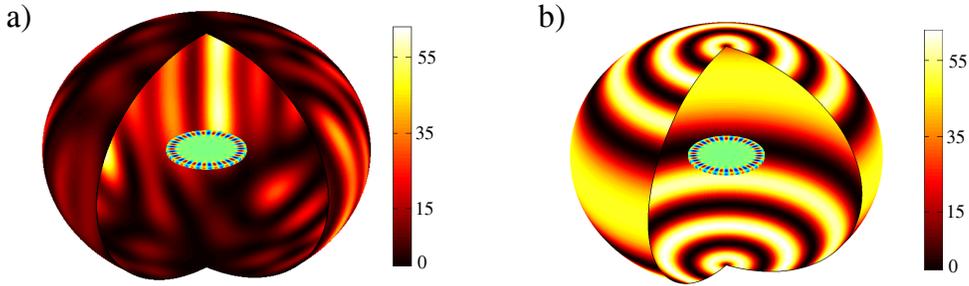

Fig. 7. Poynting vector modulus in the far field (a.u.). The calculation is made by FEM on the TE (p=1,m=21) WGM of the complete disk studied by TEM in previous section (disk 1). The disk with its WGM is visible in the middle of the sphere. (a) The complete irregular contour $\delta r(\theta)$ is employed, including the roughness and waviness extracted by the TEM analysis. (b) A simple waviness contour $\delta r(\theta) = 50\cos(21\times\theta)$ (in nm) is taken for illustrative purpose.

If the total field $\mathbf{E} = \mathbf{E^0} + \delta\mathbf{E}$ should in principle be considered to compute the Poynting vector, there are in practice some simplifications for calculating Q. Firstly, the field $\delta\mathbf{E}$ radiated by the polarization current associated to the residual roughness of the disk sums in an incoherent manner with the unperturbed field $\mathbf{E^0}$. Hence the associated "roughness" scattering losses, proportional to $Q_{rough}^{-1}$, can be strictly separated from bending losses: $Q^{-1}_{(rad+rough)} = Q_{rad}^{-1} + Q_{rough}^{-1}$. Secondly, even though the $\delta\mathbf{E}$ associated to the waviness is coherent and can interfere with $\mathbf{E^0}$, we have checked that this effect is limited in our study. It leads to about 2% error when bending losses and waviness losses are treated separately, such that we can safely use $Q^{-1}_{(rad+wav)} = Q_{rad}^{-1} + Q_{wav}^{-1}$ in what follows. Last, the separation holds true as well for the two fields $\delta\mathbf{E}$ associated to the roughness and waviness, which are respectively incoherent and coherent with $\mathbf{E^0}$, such that $Q^{-1}_{(rough+wav)} = Q_{rough}^{-1} + Q_{wav}^{-1}$. This incoherent and coherent character is directly apparent in Fig. 7. The figure shows the amplitude of the Poynting vector on the far-field sphere for an irregular disk with roughness and waviness (Fig. 7(a)) or with mere waviness (Fig. 7(b)). An incoherent "speckle-like" pattern is obtained with the rough disk, while the pattern is highly ordered when the waviness alone is considered.

Let us now apply this hybrid perturbative/numerical approach to discuss optical scattering losses in the disk analyzed in the previous section (disk1). Our TEM study has indeed shown that the contour irregularities of disk1 are representative of what is obtained by our fabrication procedure. Moreover, the parameters of its residual roughness ($\sigma$=1.02 nm and $L_c$=68 nm) are on the high side of the 5 disks studied in the TEM, thus focusing on disk1 avoids underestimating scattering losses. This disk, with a radius R=2457 nm and a thickness of 200 nm, supports four WGMs of $Q_{rad}\geq 50$ at a wavelength close to 1550 nm relevant to the experiments discussed here. These four modes are TE(p=1, m=21; $\lambda$=1537,240 nm, $Q_{rad}$=9,12.10$^9$), TE(p=2, m=17; $\lambda$=1537,003 nm, $Q_{rad}$=4,06.10$^7$), TE(p=3, m=14; $\lambda$=1525,972nm, $Q_{rad}$=4,71.10$^4$) and TE(p=4, m=11; $\lambda$=1549,996 nm, $Q_{rad}$=82). Modes of lower radiative Q, like TM modes, are disregarded. To first exemplify the role of the waviness in the optical scattering process, we inject in the model a perturbation contour $\delta r(\theta)$=50cos($m_w\times\theta$) (in nm), corresponding to a pure waviness of +-50 nm amplitude, with azimutal number $m_w$. The results of the computation of $Q_{wav}$ for the four TE modes are shown in the left panel of Fig. 8. They show that a waviness amplitude as small as 50 nm can lower $Q_{wav}$ down to below $10^3$ if the azimuthal numbers of the waviness and that of the WGM are matched. In that specific case indeed, the wavy contour acts as an efficient diffraction grating for the WGM field, thereby severely impacting the optical losses.

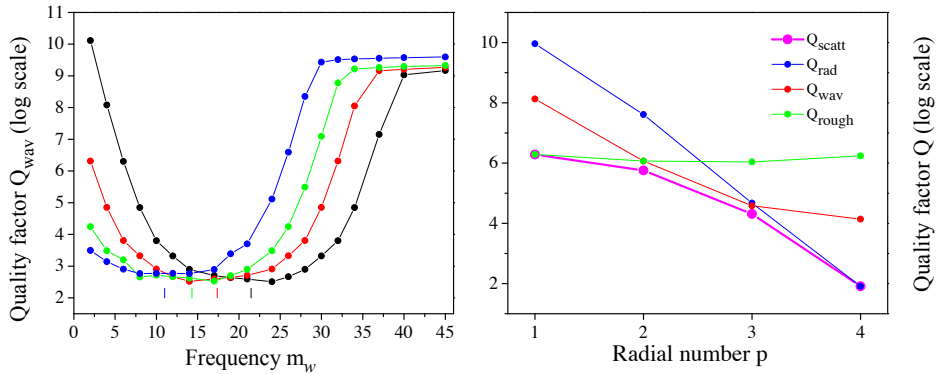

Fig. 8. Computed Q factor of the 4 TE WGMs of disk1, for various artificial or real measured contours. The calculation uses the perturbative/numerical approach discussed in the text. Left panel: An artificial wavy contour $\delta r(\theta)$=50cos($m_w\times\theta$) (in nm) is employed. The black (red, green, blue) data correspond respectively to TE (p=1,m=21; p=2,m=17; p=3,m=14; p=4,m=11). The colored vertical markers indicate the azimuthal number m of the corresponding WGM. A marked drop in $Q_{wav}$ is observed when $m_w$ approaches m. Right panel: The real contour measured by TEM on disk 1 is employed, with its different waviness components and rough residue analyzed in section 4. The fit circle of the contour is associated to $Q_{rad}$, while the waviness is associated to $Q_{wav}$ and the residual roughness to $Q_{rough}$. The total Q shown in purple is given by $Q_{scatt}^{-1}= Q_{rad}^{-1}+ Q_{wav}^{-1}+ Q_{rough}^{-1}$.

In a second stage, we now inject in the model the contour of disk 1 directly analyzed by TEM, and decompose the contributions by injecting first the waviness contour alone ($Q_{wav}$) and then the residual rough contour ($Q_{rough}$). The right panel of Fig. 8 summarizes the results obtained on the 4 TE modes that have almost equal wavelength. $Q_{wav}$ decreases with increasing radial number p. Since the wavelength is almost fixed, increasing p reduces the azimuthal number m from 21 to 11, which comes closer to the dominating azimuthal orders of the waviness of disk 1 ($1<m_w<5$). In contrast $Q_{rough}$ is almost independent of the p number. The total Q factor, accounting for the whole electromagnetic energy radiated and scattered out of the disk, varies between a few hundreds and a few millions. The picture that emerges is the following: for WGMs that are poorly confined by the curvature of the disk, such as TE (p=4, m=11) in Fig. 8, the Q factor is set by bending losses and $Q=Q_{rad}$. As the confinement progressively increases like for the TE (p=3, m=14) mode, the bending losses decrease and the wavy nature of the contour is revealed. The Q is then set by $Q^{-1}=Q_{rad}^{-1}+ Q_{wav}^{-1}$. If the WGM confinement further increases, the bending losses become negligible; the wavy losses decrease as well such that the scattering on residual roughness starts playing a role. At this stage, corresponding to TE(p=2, m=17) in Fig. 8, the Q is given by $Q^{-1}=Q_{wav}^{-1}+ Q_{rough}^{-1}$. Finally, for the strongest confinement reached by the TE(p=1, m=21) mode, the wavy losses are reduced to the point that scattering losses on residual roughness remain alone. In this last case the Q is set by roughness ($Q=Q_{rough}$) and amounts to a few millions for the disks considered in this work. In optical experiments performed on GaAs disks having the same thickness and same radius, the best Qs attain half a million [13-16], even when focusing on the p=1 or p=2 TE modes. This experimental observation also persists when scattering losses are furthered reduced by increasing the disk radius beyond the value of 2.5 µm corresponding to disk 1.

This all points towards other optical dissipation channels being responsible of the current state of the art optical Q factors of GaAs disk resonators. Scattering by contour irregularities is not the dominant loss mechanism for the best disks employed in our experiments.

## 6. Optical absorption in GaAs disks

Another loss mechanism to be considered is absorption of photons stored in the WGM. Even if our experiments are performed in a transparency region of GaAs, there are some experimental facts obviously pointing towards the existence of residual linear optical absorption. The most striking is the thermo-optic distortion of WGM resonances in the optical spectra acquired at large optical power [11,14, 33,34], see Fig. 9.

This phenomenon is explained as follows: as the laser wavelength is increased and gradually swept across a WGM resonance, photons are injected in the disk and the optical transmission drops. Residual absorption within the disk produces a local heating that increases the refractive index by means of the positive thermo-optic coefficient of GaAs. As a consequence, the WGM resonance is progressively pushed to longer wavelength as the laser continues its wavelength sweep. The cavity wavelength shift ceases when further laser wavelength increases cease to inject more light in the resonator. At this point, the optical transmission abruptly recovers the out-of resonance value. As a result the final spectrum acquires a quasi-triangular shape as observed in Fig. 9.

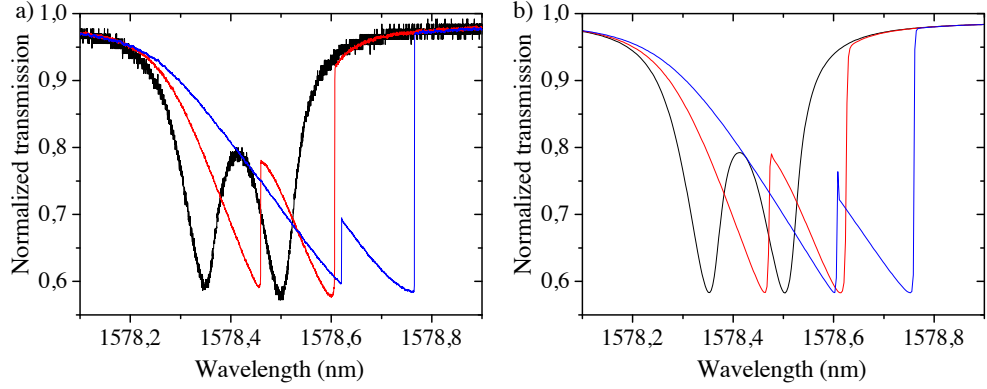

Fig. 9. Thermo-optic distortion of WGM resonances in an optical transmission spectrum. The disk radius is 2.5 µm and the thickness 200 nm (a) Experimental data. The optical power is increased from black to red to blue, making the thermo-optic triangular shape of the resonance progressively appear (15 µW, 75 µW and 150 µW respectively of output power). The employed resonance has a doublet structure due to the coupling of clockwise and counter-clockwise modes of the disk. (b) Modeling. The thermo-optic distortion is numerically modeled as explained in the text.

The thermo-optic behavior can be modeled accurately. The main thermal conduction channel between the disk and its environment is the AlGaAs pedestal (see Fig. 3(a)). The first step is to relate a temperature increase $\Delta T$ in the disk to the optical power $P_{abs}$ absorbed within the disk. Assuming a uniform temperature in the disk and a thermal gradient localized in the cylindrical pedestal, which is valid for a sufficiently narrow pedestal, we obtain the simple steady-state relation $\Delta T = h_p/(\lambda_{th}\pi r_p^2)P_{abs}$ with $\lambda_{th}$ the thermal conductivity of AlGaAs, $h_p$ and $r_p$ the height and radius of the pedestal. This relation is complemented by its dynamical version $\Delta T(t)=[\Delta T(0)-h_p/(\lambda_{th}\pi r_p^2)P_{abs}]e^{-t/\tau}+h_p/(\lambda_{th}\pi r_p^2)P_{abs}$. The second step is to relate $P_{abs}$ to the cavity frequency shift $\Delta\omega_{cav}$ induced by the temperature increase $\Delta T$. This is done by means of the coupled-mode theory, leading to the standard expression [14,16]:

$$P_{abs} = \kappa_{abs} \frac{\kappa_e}{(\omega_{cav} + \Delta\omega_{cav} - \omega)^2 + (\frac{\kappa_e}{2} + \frac{\kappa_i}{2})^2} P_{in}$$

where $\kappa_{abs}$ is the absorption rate of cavity photons, $\kappa_e$ the energy coupling rate between the cavity mode and the optical input port (the tapered fiber mode for example), $\kappa_i$ the intrinsic energy decay rate of cavity photons, $\omega_{cav}$ the cavity WGM bare frequency, $\omega$ the laser frequency and $P_{in}$ the power incident on the disk cavity. The loaded quality factor is defined by the relation $Q=\omega_{cav}/(\kappa_i+\kappa_e)$. In the limit of small $\Delta T$ valid for experiments we also have the linear relation $\Delta\omega_{cav}=(d\omega_{cav}/dn)(dn/dT)\Delta T$ with n the refractive index of GaAs, which closes a set of 3 coupled equations linking $\Delta T$, $P_{abs}$ and $\Delta\omega_{cav}$. To simulate a laser spectroscopy experiment where the laser wavelength is scanned step by step across a WGM resonance, the three equations are solved and the steady state solution found at each wavelength step serves as the starting point for the next. This iterative dynamical approach leads to the simulations shown in the right panel of Fig. 9, which satisfactorily reproduce the experimental data of the left panel. Note that most parameters in the model are fixed independently: $h_p$, $\kappa_e$, $\kappa_i$ and $P_{in}$ are measured while $(d\omega_{cav}/dn)$ and $(dn/dT)$ are known precisely. The only adjustable parameters left are $\kappa_{abs}$ and $r_p$. The pedestal radius $r_p$ is measured with limited precision in SEM images and is anyway an approximation since our pedestals are not strictly cylindrical. With this uncertainty in the effective $r_p$, the agreement with data shown in Fig. 9 indicates an absorption rate $\kappa_{abs}$ in the 1-10 GHz range, corresponding to a $Q_{abs}=\omega_{cav}/\kappa_{abs}$ between $10^5$ and $10^6$. If the precision of this evaluation is not sufficient to be perfectly conclusive, it is yet consistent with the idea that in the low optical power regime, residual linear optical absorption limits the Q of GaAs disks operated at a 1.55 µm wavelength.

At higher optical power, other nonlinear effects like two-photon absorption (TPA) are revealed. The TPA cross-section $\beta$ is defined by the propagation relation $dI/dz = -\alpha I - \beta I^2$ where I is the light intensity, $\alpha$ the linear absorption coefficient ($\alpha=\kappa_{abs}/v_g$ with $v_g$ the group velocity for the considered mode) and z the propagation direction. In bulk GaAs TPA is relatively large with $\beta$ ~10-30 cm/GW around 1.55 µm of wavelength at room temperature [35-38]. In WGM laser spectroscopy experiments, the effect of TPA would be to decrease the effective Q factor at large power. Depending if the experiments are carried-out in the under- or over-coupling regime, this should lead to a decrease or increase of the contrast of WGM resonances in the optical transmission. This behavior is indeed observed experimentally.

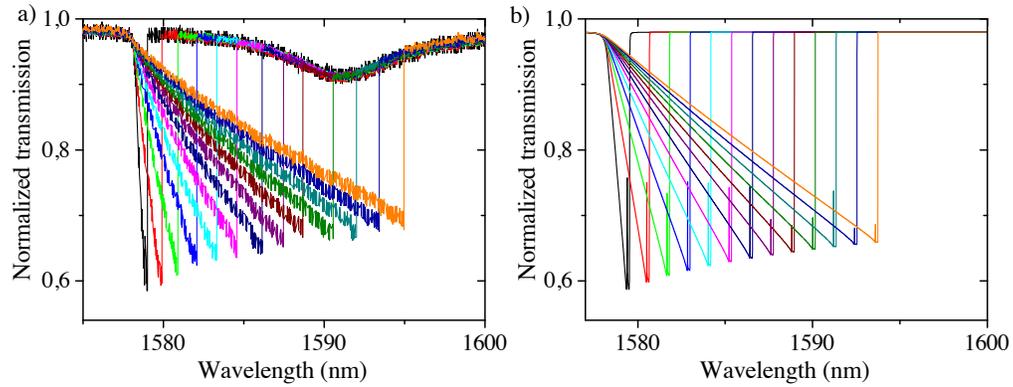

Fig. 10. Two-photon absorption in WGM spectroscopy at large optical power. (a) Experimental data. The lowest optical power (black curve) corresponds to 333 µW measured at the output of optical fiber taper, where the measured Q is $2.3\ 10^4$. The power is then multiplied by a factor 2 (red) to 13 (orange). (b) Numerical model. The behavior is reproduced by the three coupled equations discussed in the text including TPA.

Figure 10(a) shows an under-coupled WGM resonance measured at 13 distinct optical powers. As the power is increased step by step by a factor 13 in total (from black to orange), the thermo-optic distortion is accompanied by a clear reduction of the contrast. This nonlinear behavior can be modeled by completing the thermo-optic model presented above with the TPA nonlinearity. To this end, we transpose the standard $\beta$ coefficient into a TPA absorption rate $\kappa_{TPA}$ proportional to the circulating power in the cavity, which is directly injected into the coupled-mode equations describing the cavity field dynamics. In the above thermo-optic model, this results in changing $\kappa_{abs}$ into $\kappa_{abs}+\kappa_{TPA}$ and $\kappa_i$ into $\kappa_i+\kappa_{TPA}$. The results of such an approach are shown in Fig. 10(b). The agreement with experimental observations is good, with as single free parameter the effective modal volume of TPA, taken here to be an adjustable fraction of the disk volume. These results show that at the largest power employed in our experiments, multi-photon processes participate to the optical response of GaAs disks. Calculations indicate that for a disk of radius 1 µm and intrinsic optical Q of $10^5$, nonlinear losses become comparable to linear losses for a power as small as 50µW dropped into the disk.

In summary, this section comes to the conclusion that multi-photon processes dominate the optical dissipation of high-Q GaAs disks at large power, while residual linear absorption dominates at low optical power. In the final section, we experimentally study the origin of this latter undesirable dissipation channel.

## 7. Optical absorption of GaAs disks studied in liquids

If linear optical absorption below the bandgap is the limiting mechanism for high-Q GaAs cavities operated at low power, the question remains of where and why this absorption takes place. With a residual p-doping of our GaAs material below $10^{15}$ cm$^{-3}$ and statistically no dislocation present in the volume of a single disk, it is difficult to anticipate or measure the small level of bulk absorption present around 1.55 µm of wavelength. In our TEM experiments, we observed the presence of a 2 nm reconstruction layer at the surface of the disks, which may induce the presence of mid-gap states. Given the partial leakage of WGMs at the surface of the disks, and given the large surface to volume ratio of disk resonators, the surface reconstruction layer may produce substantial optical absorption below the bandgap.

To reveal and study such phenomenon, we carry out comparative spectroscopy experiments of WGMs in air, in dionized water (DI) and in diluted ammonia (NH$_4$OH 28% w/w). Liquid ammonia, like other bases and acids, is known to dissolve the reconstruction layer generally formed at the surface of GaAs in ambient conditions [39]. For these experiments, we employ GaAs disk samples of the kind introduced in [40], where GaAs tapered waveguides are suspended on chip in the disk's vicinity to allow evanescent coupling. In this configuration the disks can be easily operated in a liquid by depositing a droplet on the sample's surface. The results are summarized in Fig. 11.

In air (Fig. 11(a)), the typical thermo-optic distortion of a WGM resonance appears as the optical power is progressively increased. Just like already observed in Fig 9., the resonance has a doublet structure produced by the coupling of clockwise and counter-clockwise WGMs. When the same measurement is reproduced in DI water (Fig. 11(b)) several effects appear. First the global WGM resonance is shifted about 15 nm towards longer wavelengths. This shift results from the refractive response of water sensed by the evanescent field of the WGM, as reproduced by FEM simulations (not shown). The loaded optical Q measured at low power and the contrast of WGM resonances remain the same as in air, implying that an equivalent optical power is circulating in the disk WGM at resonance, still the thermo-optic distortion is slightly less pronounced than in air. We ascribe this difference to the added thermal conductance offered by the presence of water surrounding the disk. After drying the water droplet, the system returns quasi-reversibly to the starting situation in air. Interestingly, the behavior is quite different in diluted ammonia (Fig. 11(c)). Firstly the WGM resonance is irreversibly blue-shifted by a few nanometers in wavelength, consistent with the removal over the whole disk of the reconstruction layer of 1 to 3 nanometers of thickness. Second, the thermo-optic distortion of WGM resonances is strongly reduced, even though the loaded optical Q measured at low power and the contrast of WGM resonances are the same as in DI water, implying again an equivalent circulating in the WGM. Since ammonia and water essentially possess the same thermal properties (thermal conductivity 0.5 against 0.6 Wm$^{-1}$K$^{-1}$ and specific heat 4.7 against 4.18 kJkg$^{-1}$K$^{-1}$ at 293K and 1atm [41]) and same ability to evacuate heat, the reduction of the thermo-optic distortion points towards a reduction of the generated heat within the disk. We conclude that the in-situ removal of the reconstruction layer in liquid ammonia importantly diminishes the optical absorption in GaAs disks, which would speak for a prominent role played by surface absorption when the disks are operated in air and hence terminated by such layer. If operating resonators in ammonia reduces surface absorption, one could expect an increase of the quality factor in ammonia. In fact, water, and hence ammonia, has a large absorption coefficient in the near infrared (Fig. 11(d)), reaching the level of 13.3 cm$^{-1}$ for the wavelength of 1530 nm employed in our experiments. FEM simulations show that 15.2 % of the electromagnetic energy of the WGM locates in the liquid, resulting in an effective absorption of 2cm$^{-1}$ and a corresponding maximal Q of 6.10$^4$. This is precisely the Q value measured in the under-coupled regime in water and in ammonia, which reinforces the above interpretation.

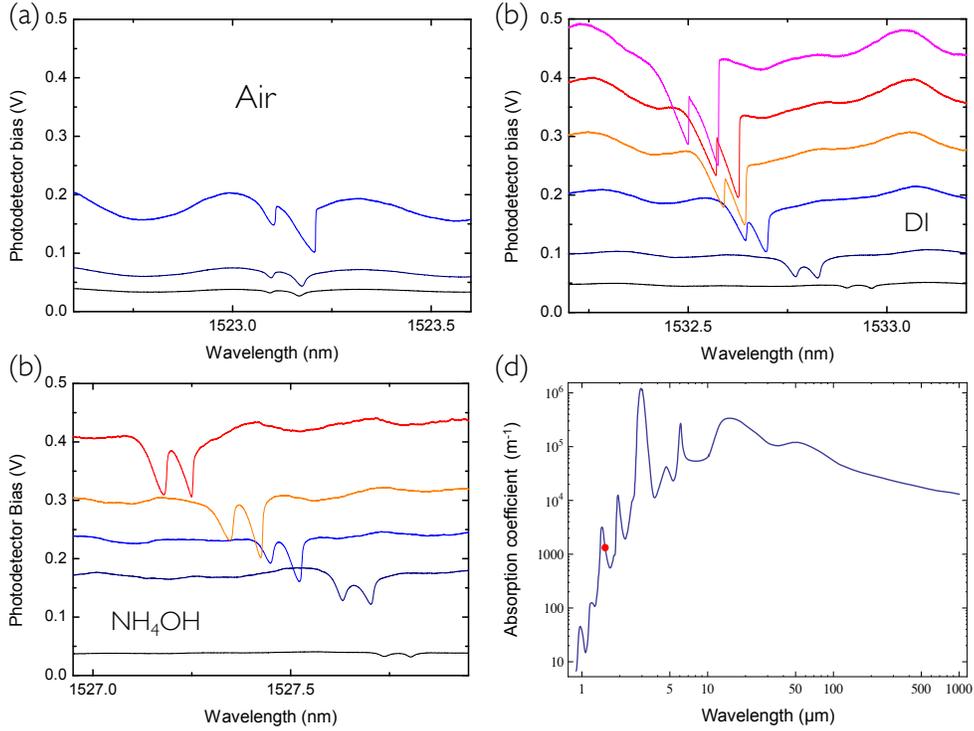

Fig. 11. WGM laser spectroscopy and thermo-optic distortion in liquids. (a) In air. The optical power is increased from black to blue, revealing the thermo-optic triangular shape of the resonance. Here again, a resonance doublet is visible because of coupling of clockwise and counter-clockwise WGMs. (b) In DI water. The optical power is increased from black to purple, with same color code as above. The thermo-optic distortion of the resonance is slightly reduced and the average wavelength red-shifted by about 15nm with respect to (a). (c) In ammonia. The optical power is increased from black to red, with same color code as above. The thermo-optic distortion is strongly reduced with respect to previous cases. (d) Water absorption spectrum shown for reference, taken from [42]. In each configuration, the out of resonance photodetector bias is proportional to the optical power $P_{in}$ circulating in the waveguide and incident onto the resonator.

## 8. Conclusions

The combination of high-resolution TEM studies, electromagnetic modeling and optical spectroscopy experiments under various conditions leads us to conclude that our best GaAs disk resonators are currently limited in their performances by residual linear absorption. The different results are consistent with a level of residual absorption corresponding to a quality factor $Q_{abs}$ between $10^5$ and $10^6$ close to 1.55 µm of wavelength. Our experiments run in liquids strongly hint towards surface absorption as being an important source of residual optical dissipation. The GaAs "native oxide" layer has a density of interface states of $10^{13}$cm$^{-2}$ located within the bandgap [39] and such density is unknown but probably sizable at the surface of our GaAs resonators. Passivation would hence be a natural way to improve their quality factors. For GaAs, treatments such as sodium sulfide (Na$_2$S), ammonium sulfide ((NH$_4$)$_2$S) [43-46] or photochemistry between GaAs and water [47] have been reported to significantly improve the rate of photoluminescence, but they produce non-permanent results. A recent passivation approach reporting permanent results over a year of time consists in forming a stable nitrogen layer at the surface of GaAs [48,49]. Such nitridation, if successfully implemented on GaAs micro and nanophotonic resonators like studied here, may lead to improvement of their Q factor. Remains to be seen which level of Q/V is attained once an efficient surface control is implemented, and what novel regimes of light-matter interaction become accessible in GaAs.

**Acknowledgments**

This work is supported by the French ANR through the NOMADE project and by the ERC through the GANOMS project.